\documentstyle[preprint,aps,epsf,floats]{revtex}

\def\overleftrightarrow{\raise1.5ex\hbox{$\leftrightarrow$}\mkern-16.5mu}

\tighten 

\tighten
\preprint{\vbox{
\hbox{DOE/ER/40762-217}
\hbox{UMD-PP-01-031}
}} \bigskip \bigskip

\begin{document}
\title{Proton-Proton Fusion in Effective Field Theory to Fifth Order}
\author{Malcolm Butler}
\address{Department of Astronomy and Physics, Saint Mary's University\\
Halifax, NS B3H 3C3 Canada\\
{\tt mbutler@ap.stmarys.ca}}
\author{Jiunn-Wei Chen}
\address{Department of Physics, University of Maryland, \\
College Park, MD 20742, USA\\
{\tt jwchen@physics.umd.edu}}
\maketitle

\begin{abstract}
The proton-proton fusion process $pp\rightarrow de^{+}\nu _{e}$ is
calculated at threshold to fifth order in pionless effective field theory.
There are two unknown two-body currents contributing at the second and
fourth orders. Combined with the previous results for $\nu _{e}d$ and $%
\overline{\nu }_{e}d$ scattering, computed to third order in the same
approach, we conclude that a $\sim $10\% measurement of reactor $\overline{%
\nu }_{e}d$ scattering measurement could constrain the $pp\rightarrow
de^{+}\nu _{e}$ rate to $\sim $7\% while a $\sim $3\% measurement of $\nu
_{e}d\rightarrow e^{-}pp$ could constrain the $pp$ rate to $\sim $2\%.
\end{abstract}

\preprint{\vbox{
\hbox{ } }} \bigskip \bigskip

\vfill\eject

The reaction $pp\rightarrow de^{+}\nu _{e}$ is of central importance to
stellar physics and neutrino astrophysics. Bethe and Critchfield proposed it
to be the first reaction to ignite the $pp$ chain nuclear reactions that
provided the principal energy and neutrinos in the Sun \cite{Bet38}. Major
efforts have been made to provide precise theoretical predictions for this
reaction at zero energy \cite
{Sal52,Bli65,Bah69,Gar72,Dau76,Bar79,Gou90,Car91,Kam94,Iva97,Sch98,Par98}.
No direct experimental constraint is available for this process, and the
accuracy of the theoretical predictions must always be weighed against the
availability of empirical constraints. However, Schiavilla {\it et al.}\/
recently calibrated the matrix element with complementary calculations of
tritium beta decay to obtain an estimated uncertainty of less than 1\% in
the potential model.

Alternatively, effective field theory (EFT) can provide a connection between 
$pp$ fusion and other processes that might be accessible experimentally.
Specifically, there is a direct connection between the reactions $%
pp\rightarrow de^{+}\nu _{e}$ and $\nu _{e}d\rightarrow e^{-}pp$ in that
they both involve the same matrix elements. Further, in EFT, they can both
be shown to depend on a single unknown counterterm (or two-body current) at
third order. Otherwise, the nuclear physics input is identical. Further, all
four $\nu (\overline{\nu })$-$d$ breakup channels depend on this same
counterterm, meaning that a measurement in any one channel implies a
measurement in all channels -- including $pp$ fusion. To date, measurements
are available for $\overline{\nu }_{e}d$ breakup using reactor antineutrinos
with $\sim 10$-$20\%$ uncertainty \cite{Pas79,Vid90,Ver91,Ril99,Koz99} and
possibly $\sim 5\%$ in the future \cite{Koz99}. Also, the ORLaND proposal
could provide a measurement of $\nu _{e}d\rightarrow e^{-}pp$ to a few
percent level \cite{ORLaND}. It is important to understand the constraints
that these measurements could on $pp\rightarrow de^{+}\nu _{e}.$

At the present time, $\nu _{e}d$ and $\overline{\nu }_{e}d$ breakup
processes have been studied to the third order in pionless EFT depending on
the unknown two-body counterterm $L_{1,A}$. Through varying $L_{1,A}$, the
four channels of potential model results of refs.\cite{YHH} and \cite{NSGK}
are reproduced to high accuracy. This confirms that the $\sim 5\%$
difference between refs.\cite{YHH} and \cite{NSGK} is largely due to
different assumptions made to short distance physics. We will discuss this
difference later in this letter.

For $pp$ fusion, an\ pionless EFT calculation has been performed to second
order by Kong and Ravndall\cite{KRpp}, with a resulting dependence on $%
L_{1,A}$. In this letter, we use the same approach to push the calculation
to fifth order, which introduced another unknown two-body counterterm ($%
K_{1,A}$) that first enters at fourth order. Constraining $K_{1,A}$ using
dimensional analysis, we conclude that a measurement of reactor $\overline{%
\nu }_{e}d$ with 10\% uncertainty can constrain $pp$ fusion to 7\% while a
measurement of $\nu _{e}d\rightarrow e^{-}pp$ to 3\% would constrain $pp$
fusion to 2\%.

The method we use is pionless nuclear effective field theory, ${\rm EFT}({%
\pi \hskip-0.6em/})$ \cite{CRS}, treating the electromagnetic interaction
between protons in the manner developed in \cite{KRpp,KRem}. The dynamical
degrees of freedom are nucleons and non-hadronic external currents. Massive
hadronic excitations such as pions and the delta are integrated out,
resulting in contact interactions between nucleons. The nucleons are
non-relativistic but with relativistic corrections built in systematically.
Nucleon-nucleon interactions are calculated perturbatively with the small
expansion parameter 
\begin{equation}
Q\equiv \frac{(1/a_{pp},\gamma ,p,\alpha M_{N})}{\widetilde{\Lambda }}
\end{equation}
which is the ratio of the light to heavy scales. The light scales include
the inverse S-wave nucleon-nucleon scattering length $1/a_{pp}(a_{pp}=-7.82$
fm$)$ in the $^{1}S_{0}$, $pp$ channel, the deuteron binding momentum $%
\gamma (=45.69$ MeV) in the $^{3}S_{1}$ channel, the proton momentum in the
center-of-mass frame $p$, and the fine structure constant $\alpha (=1/137)$
times the nucleon mass $M_{N}$. The heavy scale $\widetilde{\Lambda }$,
which dictates the scales of effective range and shape parameters is set by
the pion mass $m_{\pi }$. The Kaplan-Savage-Wise renormalization scheme \cite
{KSW} is used to make the power counting \cite{KSW,vK97} in $Q$ transparent.
This formalism has been successfully applied to many processes involving the
deuteron\cite{CRS,npdgam2} including Compton scattering \cite{dEFT,GR}, $%
np\rightarrow d\gamma $ for big-bang nucleosynthesis \cite{npdgam1,Rupak}, $%
\nu d$ scattering \cite{BCK} for physics at the Sudbury Neutrino Observatory 
\cite{SNO}, and parity violation observables \cite{PV}.

There are other power counting schemes which yield different orderings in
the perturbative series and each has certain advantages. For example, the
z-parametrization \cite{Zpara} recovers the exact deuteron wave function
renormalization at second order; the dibaryon pionless EFT \cite{dEFT}
resums the effective range parameter contributions at first order and
simplifies the calculation tremendously by cleverly employing the equations
of motion to remove redundancies in the theory. While one of these power
countings could lead to more rapid convergence in any given calculation, for
a high-order calculation as we present here the distinctions between
different expansions are negligible.

Ignoring for the moment the weak interaction component, the relevant
lagrangian in ${\rm EFT}({\pi \hskip-0.6em/})$ can be written as a
derivative expansion 
\begin{eqnarray}
{\cal L} &=&N^{\dagger }\left( i\partial _{0}+{\frac{{\bf \nabla }^{2}}{%
2M_{N}}}\right) N  \nonumber \\
&&-C_{0}^{\left( ^{3}S_{1}\right) }(N^{T}P_{i}N)^{\dagger }(N^{T}P_{i}N) 
\nonumber \\
&&+{\frac{C_{2}^{\left( ^{3}S_{1}\right) }}{8}}\left[ (N^{T}P_{i}N)^{\dagger
}(N^{T}\raise1.5ex\hbox{$\leftrightarrow$}\mkern-16.5mu{\bf \nabla }%
^{2}P_{i}N)+h.c.\right]  \nonumber \\
&&-{\frac{C_{4}^{\left( ^{3}S_{1}\right) }}{16}}(N^{T}\raise1.5ex%
\hbox{$\leftrightarrow$}\mkern-16.5mu{\bf \nabla }^{2}P_{i}N)^{\dagger
}(N^{T}\raise1.5ex\hbox{$\leftrightarrow$}\mkern-16.5mu{\bf \nabla }%
^{2}P_{i}N)\quad  \nonumber \\
&&-{\frac{\widetilde{C}_{4}^{\left( ^{3}S_{1}\right) }}{32}}\left[ (N^{T}%
\raise1.5ex\hbox{$\leftrightarrow$}\mkern-16.5mu{\bf \nabla }%
^{4}P_{i}N)^{\dagger }(N^{T}P_{i}N)+h.c.\right]  \nonumber \\
&&+\left( ^{3}S_{1}\rightarrow \,^{1}S_{0},\ P_{i}\rightarrow \overline{P}%
_{i}\right) +\cdots \ ,
\end{eqnarray}
where $\raise1.5ex\hbox{$\leftrightarrow$}\mkern-16.5mu{\bf \nabla }\equiv 
\overrightarrow{{\bf \nabla }}-\overleftarrow{{\bf \nabla }}$ and where $%
P_{i}=\sigma _{2}\sigma _{i}\tau _{2}/\sqrt{8}$ and $\overline{P}_{i}=\sigma
_{2}\tau _{2}\tau _{i}/\sqrt{8}$ project out $^{3}S_{1}$ and $^{1}S_{0}$
channels respectively with $\sigma (\tau )$ acting on spin(isospin) indices.
The coupling constants have been fit to data in \cite{CRS,BCK}. We only
perform the calculation at threshold and thus the relativistic corrections
and momentum transfer (${\bf q}$) effects are suppressed by factors of $%
\gamma ^{2}/M_{N}^{2}$ and ${\bf q}^{2}/\gamma ^{2}$ , both of which are $%
\ll 1\%$. Thus, with the goal of presenting a calculation with a precision
of less than 1\%, the nonrelativistic and zero recoil limits are suitable
approximations for us to make.

The weak interaction terms in the lagrangian can be written as 
\begin{equation}
{\cal L}_{W}=-\frac{G_{F}}{\sqrt{2}}l_{+}^{\mu }J_{\mu }^{-}+h.c.+\cdots \ ,
\end{equation}
where $G_{F}$ is the Fermi decay constant and $l_{+}^{\mu }=\overline{\nu }%
\gamma ^{\mu }(1-\gamma _{5})e$ is the leptonic current. The hadronic
current has vector and axial vector parts, 
\[
J_{\mu }^{-}=V_{\mu }^{-}-A_{\mu }^{-}\ . 
\]
The vector current matrix element vanishes in the zero recoil limit since it
yields terms proportional to the wave function overlapping between two
orthogonal states. The time component of the axial current gives center of
mass motion corrections which vanish in our approximation. The spatial
component of the axial vector is the dominant contribution, and can be
written in terms of one-body and two-body currents 
\begin{eqnarray}
A_{k}^{-} &=&{\frac{g_{A}}{2}}N^{\dagger }\tau ^{-}\sigma _{k}N  \nonumber \\
&&+L_{1,A}\left[ \left( N^{T}P_{k}N\right) ^{\dagger }\left( N^{T}\overline{P%
}^{-}N\right) +h.c.\right]  \nonumber \\
&&+\frac{K_{1,A}}{8}\left[ \left( N^{T}\raise1.5ex\hbox{$\leftrightarrow$}%
\mkern-16.5mu{\bf \nabla }^{2}P_{k}N\right) ^{\dagger }\left( N^{T}\overline{%
P}^{-}N\right) +\left( N^{T}P_{k}N\right) ^{\dagger }\left( N^{T}\raise1.5ex%
\hbox{$\leftrightarrow$}\mkern-16.5mu{\bf \nabla }^{2}\overline{P}%
^{-}N\right) +h.c.\right]  \nonumber \\
&&+\cdots \ ,
\end{eqnarray}
where $g_{A}=1.26$, $\tau ^{-}=(\tau _{1}-i\tau _{2})$, and the values of $%
L_{1,A}$ and $K_{1,A}$ are yet to be determined by data.

The hadronic matrix element is usually parametrized in the form 
\begin{equation}
\left| \left\langle d;j\left| A_{k}^{-}\right| pp\right\rangle \right|
\equiv g_{A}C_{\eta }\sqrt{\frac{32\pi }{\gamma ^{3}}}\Lambda (p)\delta
_{k}^{j}\ ,
\end{equation}
where $j$ is the deuteron polarization state, and 
\begin{equation}
C_{\eta }=\sqrt{\frac{2\pi \eta }{e^{2\pi \eta }-1\ }}\ ,\quad \eta =\frac{%
\alpha M_{N}}{2p}
\end{equation}
is the well-known Sommerfeld factor. In the center of the Sun, the scale of $%
p$ is $\sim $1 MeV. Thus the Sommerfeld factor $C_{\eta }^{2}$ changes
rapidly with respect to $p$ while $\Lambda (p)$ does not. It is sufficient
to keep the $p^{2}$ correction for $\Lambda (p)$, since the higher order
correction is ${\cal O}($ $\left( p/\alpha M_{N}\right) ^{4})$. Using $%
\epsilon $ to keep track of the $Q$ expansion we find that, to fifth order
in the $Q(\epsilon )$ expansion, $\Lambda (0)$ can be written in the compact
form 
\begin{equation}
\Lambda (0)=%
{\displaystyle{1 \over \sqrt{1-\epsilon \gamma \rho _{d}}}}%
\left\{ e^{\chi }-\gamma a_{pp}\left[ 1-\chi e^{\chi }E_{1}(\chi )\right]
-\epsilon \gamma ^{2}a_{pp}\left[ \overline{L}_{1,A}-\frac{\epsilon
^{2}\gamma ^{2}}{2}\overline{K}_{1,A}\right] \right\} +{\cal O}\left(
\epsilon ^{5}\right) \ .  \label{Lambda0}
\end{equation}
This expression can be expanded to $\epsilon ^{4}$, and then $\epsilon $
should be set to 1. $\rho _{d}=1.764$ fm is the effective range parameter in
the $^{3}S_{1}$ channel, $\chi =\alpha M_{N}/\gamma $ and 
\begin{equation}
E_{1}(\chi )=\int_{\chi }^{\infty }dt\frac{e^{-t}}{t}\ .
\end{equation}
$\overline{L}_{1,A}$ and $\overline{K}_{1,A}$ are the renormalization scale $%
\mu $-independent combinations of the $\mu $-dependent parameters $L_{1,A}$, 
$K_{1,A}$ and the nucleon-nucleon contact terms $C_{2}$ and $C_{4}$ 
\begin{eqnarray}
\overline{L}_{1,A} &=&\frac{-\left( \mu -\gamma \right) }{%
M_{N}C_{0,-1}^{(pp)}}\left[ \frac{L_{1,A}}{g_{A}}-\frac{M_{N}}{2}\left(
C_{2,-2}^{(pp)}+C_{2,-2}^{(d)}\right) \right] \ ,  \nonumber \\
\overline{K}_{1,A} &=&\frac{-\left( \mu -\gamma \right) }{%
M_{N}C_{0,-1}^{(pp)}}\left[ \frac{K_{1,A}}{g_{A}}-M_{N}\left( \widetilde{C}%
_{4,-2}^{(pp)}+2C_{4,-3}^{(d)}\right) \right] \ .
\end{eqnarray}
>From ref. \cite{BCK}, we have 
\begin{eqnarray}
C_{0,-1}^{(pp)} &=&%
{\displaystyle{4\pi  \over M_{N}}}%
{\displaystyle{1 \over %
{\displaystyle{1 \over a_{pp}}}-\mu +\alpha M_{N}\left( \ln %
{\displaystyle{\mu \sqrt{\pi } \over \alpha M_{N}}}+1-%
{\displaystyle{3 \over 2}}\gamma _{E}\right) }}%
\ ,  \nonumber \\
C_{2,-2}^{(pp)} &=&%
{\displaystyle{M_{N} \over 8\pi }}%
r_{0}^{(pp)}C_{0,-1}^{(pp)}{}^{2}\ ,\quad \ \widetilde{C}_{4,-2}^{(pp)}=-%
{\displaystyle{M_{N} \over 4\pi }}%
r_{1}^{(pp)}C_{0,-1}^{(pp)}{}^{2}\ ,  \nonumber \\
C_{2,-2}^{(d)} &=&%
{\displaystyle{2\pi  \over M_{N}}}%
\frac{\rho _{d}}{\left( \mu -\gamma \right) ^{2}}\ ,\ \quad C_{4,-3}^{(d)}=-%
{\displaystyle{\pi  \over M_{N}}}%
\frac{\rho _{d}^{2}}{\left( \mu -\gamma \right) ^{3}}\ ,
\end{eqnarray}
where the second subscripts of the $C$'s denote the scaling in powers of $Q$%
, $\gamma _{E}=0.577$ is Euler's constant, and $r_{0}^{(pp)}=2.79$~fm is the 
$pp$ channel effective range. We take the $pp$ channel shape parameter to be
the same as that in the $np$ channel $r_{1}^{(pp)}=-0.48$ fm$^{3}$. Any
errors introduced by this are small numerically.

After expanding to $\epsilon ^{4}$ and setting $\mu =m_{\pi }$, we obtain 
\begin{equation}
\Lambda (0)=2.58+0.011\left( \frac{L_{1,A}}{\text{1 fm}^{3}}\right)
-0.0003\left( \frac{K_{1,A}}{\text{1 fm}^{5}}\right) \ .  \label{Lambda0N}
\end{equation}
At $\mu =m_{\pi }$, dimensional analysis as developed in refs. \cite{CRS,KSW}
would favour 
\begin{eqnarray}
\left| L_{1,A}\right| &\approx &\frac{1}{m_{\pi }\left( m_{\pi }-\gamma
\right) ^{2}}\approx 6\text{ fm}^{3}\ ,  \nonumber \\
\left| K_{1,A}\right| &\approx &\frac{1}{m_{\pi }^{2}\left( m_{\pi }-\gamma
\right) ^{3}}\approx 20\text{ fm}^{5}\ .  \label{da}
\end{eqnarray}
Two observations follow. First, if we take these naively estimated values
(with positive signs, for example), then the expansion of $\Lambda (0)$
converges rapidly 
\begin{equation}
\Lambda (0)=2.51\left( 1+0.039+0.029-0.010-0.0001\right) \ .
\end{equation}
For other sign combinations, the series also converges rapidly and higher
order effects are $\ll 1\%$. Second, eqs. (\ref{Lambda0N}) and (\ref{da})
show that $K_{1,A}$ is likely to contribute to $\Lambda (0)$ at a level less
than 1\%. Thus eq. (\ref{Lambda0N}) is precise to 1\% even with $K_{1,A}$
set to zero, meaning that we can write 
\begin{equation}
\Lambda (0)=2.58+0.011\left( \frac{L_{1,A}}{\text{1 fm}^{3}}\right) +{\cal O}%
(1\%)\ .  \label{central}
\end{equation}
This the central result of this paper.

Ultimately the value of $L_{1,A}$ must be extracted from experimental data.
Before we address this issue, let us look at what values of $L_{1,A}$ are
found in fits to potential model calculations. Using the results of ref. 
\cite{BCK}, the recent potential model results for $\nu (\overline{\nu })-d$
breakup of Nakamura, Sato, Gudkov, and Kubodera (NSGK) \cite{NSGK} are
equivalent to 
\begin{equation}
L_{1,A}^{NSGK}=5.6\text{ }\pm 2\ \text{fm}^{3}\ ,
\end{equation}
while the results of Ying, Haxton and Henley (YHH) \cite{YHH} are equivalent
to 
\begin{equation}
L_{1,A}^{YHH}=0.94\pm 2\ \text{fm}^{3}\ .
\end{equation}
The uncertainties represent a conservative estimate from EFT of 3\% from
ref.~\cite{BCK} (even though the NSGK results can be reproduced within 1\%
using the central value of $L_{1,A}$). Based on the size of the third-order
contribution, the actual uncertainty may be as small as 1\%, but further
analysis is required to ascertain this. The (undefinable) errors from the
potential models themselves are not included. As mentioned before, the
differences between these two calculations are in their treatment of
two-body physics. NSGK uses a model to include axial two-body meson exchange
currents, while YHH includes vector two-body currents but not the
more-important axial two-body currents. Given that $L_{1,A}$ is
representative of the dominant axial effects in EFT, it is not surprising to
see substantial differences in the value inferred by each calculation.

Similarly, eq. (\ref{central}) translates the $pp$ fusion rate ($\Lambda
^{2}(0)=7.05$-$7.06$) calculated by Schiavilla {\em et al.} into 
\begin{equation}
L_{1,A}^{Schiavilla\ et\ al.}=6.5\pm 2.4\text{ fm}^{3}\ ,
\end{equation}
which is consistent with NSGK.

Experimentally, it is easy to relate the EFT $\nu (\overline{\nu })$-$d$
scattering results and $pp$ fusion rate through the third-order results for $%
\nu (\overline{\nu })$-$d$ provided in \cite{BCK}. The results for each
channel are parameterized in the form 
\begin{equation}
\sigma (E_{\nu })=a(E_{\nu })+b(E_{\nu })L_{1,A}+{\cal O}(< 3\%)\ ,
\end{equation}
where $a$ and $b$ are functions of neutrino energy. These results, tabulated
in ref.~\cite{BCK}, can be easily related to the $pp$ fusion rate through
eq. (\ref{central}).

For reactor $\overline{\nu }$-$d$ scattering, one expects the rate to be
peaked around 8 MeV, which can be interpreted to mean a measurement precise
to 10\% can determine the $pp$ fusion rate ($\propto \Lambda ^{2}$) to 7\%.
With the experimental precision further improved to 5\%, the $pp$ fusion
rate can be determined to 4\%. Alternatively, the proposed ORLaND \cite
{ORLaND} detector might measure $\nu _{e}d\rightarrow e^{-}pp$ to 3\%, in
turn constraining the $pp$ fusion rate to 2\%.

In strong interaction processes involving external currents, delicate
relations between operators are required to guarantee that there are no off-shell
ambiguities in the final results. For example, in two-body systems two-body
currents serve to absorb the off-shell effects from the two-body strong
interactions. This means that different models that reproduce the same on-shell
nucleon-nucleon scattering data might need quite distinct two-body
currents to deal with off-shell effects. In the case of $pp$ fusion or $\nu $-$d$ scattering,
electromagnetic matrix elements cannot be used to constrain weak matrix elements.
Their operator structures are quite different at the quark level even
they might appear the same in hadronic level. Using tritium $\beta $-decay
to constrain the two-body current \cite{Sch98} is an excellent idea, in
principle. However, contributions from three-body currents that are
required to absorbed three-body off-shell effects are not yet
constrained. In light of these facts, the need for a precise experimental
measurement of $\nu _{e}(\overline{\nu }_{e})$-$d$ breakup cannot be
overstated.




\vskip2in \centerline{\bf ACKNOWLEDGMENTS} We would like to thank Ian Towner
for useful discussions. M.B.\ is supported by a grant from the Natural
Sciences and Engineering Research Council of Canada. J.-W.C.\ is supported,
in part, by the Department of Energy under grant DOE/ER/40762-213.

\end{document}